\documentclass[aps,prd,twocolumn,superscriptaddress,aas_macros,nofootinbib,10pt]{revtex4-1}
\usepackage{graphicx}
\usepackage{subfigure}
\usepackage{amsmath} 

\usepackage[colorlinks=true,citecolor=blue,urlcolor=magenta,breaklinks]{hyperref}
\usepackage{natbib}
\usepackage{newtxtext,newtxmath}
\usepackage[T1]{fontenc} 
\begin{document}

\title[Dark matter imprint  on $^8$B neutrino spectrum]{
Dark matter imprint  on $^8$B  neutrino spectrum}

\author{Il\'idio Lopes}
\email[]{ilidio.lopes@tecnico.ulisboa.pt}
\affiliation{Centro de Astrof\'{\i}sica e Gravita\c c\~ao  - CENTRA, 
 Departamento de F\'{\i}sica, Instituto Superior T\'ecnico - IST,
 Universidade de Lisboa - UL, Av. Rovisco Pais 1, 1049-001 Lisboa, Portugal}
 \author{Joseph Silk}
\email[]{silk@astro.ox.ac.uk}
\affiliation{Institut d'Astrophysique de Paris, UMR 7095 CNRS, Universit\'e Pierre et Marie Curie, \\ 98 bis Boulevard Arago, Paris 75014, France}
\affiliation{Beecroft Institute of Particle Astrophysics and Cosmology, 1 Keble Road, University of Oxford, Oxford OX1 3RH, United Kingdom}
 \affiliation{Department of Physics and Astronomy, 3701 San Martin Drive, The Johns Hopkins University, Baltimore Maryland 21218, USA}

% % % % % % % % % % % % % % % % % % % % % % % % % % % % % % % % % % % % % % %
\date{\today}

\begin{abstract} 
The next generation of solar neutrino detectors will provide a precision measure of the $^8$B electron-neutrino spectrum in the energy range from 1-15 MeV. 
Although the neutrino spectrum emitted by $^8$B $\beta$-decay reactions in the Sun's core is identical to the neutrino spectrum  measured in the laboratory,  due to vacuum and   matter  flavor oscillations,  this spectrum  will be very different from  that  measured on Earth by the different solar neutrino experiments.  We study how the presence of dark matter (DM) in the Sun's core changes the shape of the $^8$B electron-neutrino spectrum. These modifications  are caused by local variations of the electronic density and the $^8$B neutrino source, induced by local changes of the temperature, density and chemical composition.  
Particularly  relevant are the shape changes at low and medium energy range ($E_\nu\le 10 {\; \rm MeV}$),  for which the experimental noise level is expected to be quite small. If such a distortion in the $^8$B$\nu_e$ spectrum were to be observed, this would strongly hint in favor  of the existence of DM in the Sun's core. The $^8$B electron-neutrino spectrum provides a  complementary method to helioseismology and total neutrino fluxes for constraining the DM properties. In particular, we study the impact of light asymmetric DM on solar neutrino spectra. Accurate neutrino spectra measurements could help to determine whether light asymmetric DM exists in the Sun's core, since it has been recently advocated
that this type of DM  might resolve the solar abundance problem.  
\end{abstract}

%\pacs{ln}
%\keywords{ln}


\maketitle
% 
%%%%%%%%%%%%%%%%%%%%%%%%%%%%%%%%%%%%%%%%%%%%%%%%%%%%%%%%%%%%%%%%%%%%%%%%%%%%%
%
%%%%%%%%%%%%%%%%%%%%%%%%%%%%%%%%%%%%%%%%%%%%%%%%%%%%%%%%%%%%%%%%%%%%%%%%%%%%%
\section{Introduction}
%%%%%%%%%%%%%%%%%%%%%%%%%%%%%%%%%%%%%%%%%%%%%%%%%%%%%%%%%%%%%%%%%%%%%%%%%%%%%
Solar neutrino detectors have been one of the beacons of particle physics, 
both by leading the way in discovering the basic properties of particles, 
including the nature of neutrino flavor oscillations, and by being responsible 
for developing pioneering techniques in experimental neutrino detection~\citep[e.g.,][]{2012APh....35..685W,2013ARA&A..51...21H,2018PrPNP..98....1G}. 
The next generation of  detectors like the DUNE Experiment~\citep{2016NuPhB.908..318D}, the CJPL Laboratory~\citep{2017ChPhC..41b3002B}, the JUNO  Observatory~\citep{2016JPhG...43c0401A}, and the LENA detector~\citep{2012APh....35..685W},
will measure with high precision the neutrino fluxes  and neutrino spectra of a few key solar nuclear reactions, such as the electron-neutrino ($^8$B$\nu_e$)  spectrum produced by the  $\beta$-decay of $^8$B
 ~\citep{2009RPPh...72j6201B,2013arXiv1310.4340D}.   This will allow us to probe in detail the Sun's core, including the search for unknown physics processes. 
Moreover, the high quality of the data will enable the development of  inversion techniques for determining 
basic properties of the solar plasma~\citep[e.g.,][]{2005NuPhS.138..347B}. 
Specific examples can be found in~\citet{1998PhLB..427..317B},
~\citet{2016PhRvL.117u1101D} and~\citet{2013ApJ...777L...7L}. 
Equally, solar neutrino data can be used to find specific features associated with new physical processes~\citep[e.g.][]{2013PhRvD..87d3001S,2017PhRvD..95a5023L}, 
such as the possibility of an isothermal solar core associated with the presence of
DM~\citep{2002PhRvL..88o1303L}.

The $^8$B$\nu_e$ spectrum emitted by the nuclear reactions in the  
Sun's core is identical to that determined  by current laboratory experiments~\citep[e.g.][]{2000PhRvL..85.2909O,2003PhRvL..91y2501W,2006PhRvC..73b5503W,2011PhRvC..83f5802K,2012PhRvL.108p2502R}.
\citet{1991PhRvD..44.1644B} has shown that the 
corrections on  the shape of neutrino energy spectra caused by the surrounding plasma 
in the Sun's core are negligible. For example,  the corrections related with the thermal 
motions of the colliding ions are negligible, 
as the thermal velocity of ions is much smaller than the velocity of light. 
Given that the $^8$B$\nu_e$ spectrum 
shape is well-known and we know that the solar plasma does not influence 
significantly the nuclear reactions occurring in the Sun's core, the changes detected 
in the $^8$B$\nu_e$ spectrum will be mostly due to neutrino flavor oscillations induced by matter~\citep[a process also known as the Mikheyev-Smirnov-Wolfenstein: MSW effect,][]{1958JETP....6..429P,1978PhRvD..17.2369W,2013PhRvD..88d5006L}. 
These modifications will change not only the overall neutrino flux but, more significantly, modify
the neutrino spectrum by affecting in a differential manner the survival probability of electron-neutrinos -- depending upon the energy of the emitted neutrino. 

The impact of neutrino flavor oscillations on the total neutrino fluxes is extensively documented 
in the literature~\citep[see ][and references therein]{2013ARA&A..51...21H}, but the impact on solar neutrino  spectra has been discussed only briefly.
The reason is that  neutrino flavor oscillations are expected to be unimportant, because the temperature of the Sun's core is strongly constrained by the total neutrino fluxes. Nevertheless, as we discuss in this article, this is not necessarily the case, mostly because neutrino flavor oscillations due to the MSW effect in the Sun's core depend strongly of the local properties of the solar plasma. Unlike total neutrino fluxes, these give  differential information about the physics
of the Sun's core.  In particular, if light DM is present in the solar core,  the amount of electron-neutrinos converted to other 
flavors will be different from the value found in the standard solar model 
~\citep[SSM, e.g.,][]{2010ApJ...713.1108G,2011ApJ...743...24S}, and consequently their total neutrino fluxes and  neutrino spectra will also be different from the SSM.  
In recent years,  significant improvements in the measurement accuracy of solar neutrino fluxes  have been instrumental in allowing the use of the Sun to  set constraints on  the properties of dark matter, including the neutralino~\citep[e.g.,][]{2002PhRvD..66e3005B,2012ApJ...746L..12T}.
and impose limits to the expected neutrino fluxes coming from the Sun due to  DM annihilation~\citep[e.g.,][]{2016ApJ...827..130L}.  Moreover, a large number of  different types of asymmetric DM have been discussed in the literature~\citep{2010PhRvD..82j3503C,2010PhRvL.105a1301F,2010PhRvD..82h3509T,2012PhRvL.108f1301I,2012ApJ...757..130L}. 
The presence 
of dark matter in the Sun's core could help solve the long-running
solar composition problem~\citep{2009ApJ...705L.123S}, a discrepancy between the solar structure inferred from helioseismology and the one computed from a SSM by inputting the most up-to-date photospheric abundances~\citep{2014ApJ...795..162L,2015JCAP...08..040V}.
This type of diagnostic has also been successfully extended to other stars, including other  sun-like stars~\citep[e.g.,][]{2013ApJ...765L..21C,2017PhRvD..95b3507M}
 and neutron stars~\citep{2010PhRvD..82f3531K,2011PhRvL.107i1301K,2011PhRvD..83h3512K,2014AIPC.1604..389K}. 
In addition, such types of studies have also been extended to the  first generation of stars~\citep{2008PhRvD..78l3510T,2009MNRAS.394...82S,2011ApJ...733L..51C,2011ApJ...742..129S,2014ApJ...786...25L}.

In this paper, we show that by measuring the  $^8$B$\nu_e$ solar spectrum, 
it is possible to constrain the DM content in the Sun's core.  
This diagnostic complements the total neutrino flux analysis.
This is a robust result, as the shape variation of the  $^8$B$\nu_e$ spectrum is uniquely  related to the radial variation of the plasma properties in the Sun's core, where the maximum accumulation of DM is expected to 
occur. This type of diagnostic is particularly useful for testing new types of DM models~\citep[e.g.,][]{2014ApJ...780L..15L,2014JCAP...04..019V},  
which have a more pronounced impact in the core of the Sun.

%%%%%%%%%%%%%%%%%%%%%%%%%%%%%%%%%%%%%%%%%%%%%%%%%%%%%%%%%%%%%%%%%%%%%%%%%%%%%
\section{Current status of Dark matter research}
%%%%%%%%%%%%%%%%%%%%%%%%%%%%%%%%%%%%%%%%%%%%%%%%%%%%%%%%%%%%%%%%%%%%%%%%%%%%%
%
	In the last few years, several types of light DM particles have been suggested as an  ideal DM candidate  for a elementary DM particle, motivated by fundamental theoretical arguments in cosmology and particle physics, 
	and by a few positive hints from some direct DM search experiments. 
	Nevertheless, these results are controversial since other experimental detectors have excluded the same DM parameter space. 
\smallskip

In favor of the theoretical argument, several DM models 
succeed in explaining the observed DM relic density~\citep{2013ApJS..208...20B,2014A&A...571A..16P}, 
as a new type of light DM, usually referred to as asymmetric DM~\citep{2013IJMPA..2830028P,2014PhR...537...91Z,2014AIPC.1604..389K}.
Unlike symmetric DM, this new type of DM is believed to be produced 
in the primordial universe by physical mechanisms identical to the production of baryons, 
known as darkogenesis~\citep[e.g.,][]{2011PhLB..699..364D,2011PhRvD..83e5008G,2009NuPhB.812..243C,2010PhLB..687..275D}, 
and likewise composed of an unbalanced mixture of particles and antiparticles. 
The proportionality of DM particles relatively to DM antiparticles is measured by 
the asymmetry parameter $\eta_{\rm DM}$, which is identical to $\eta_{\rm B}$ for baryons. 
In the case  that the DM is symmetric, i.e., the DM particle is its own antiparticle, $\eta_{\rm DM}=0$.

The production of asymmetric DM in the early universe is computed by a similar
procedure to baryogenesis~\citep[e.g.][]{2011JCAP...07..003I,2006PhRvD..73l3502D}.
The origin of such a DM asymmetry is not known; however, as suggested by some extensions to the standard model 
of particle physics, this could be related to the
existence of electric and magnetic dipole moments of some standard particles and possibly  new particles~\citep[e.g.,][]{2004PhRvD..70h3501S}. For reference, the current upper limit on the electric 
dipole moment of the electron is set to $8.7\times 10^{-29} {\rm e \; cm }$~\citep{2014Sci...343..269B}. 
Similarly~\citet[][]{2014ApJ...780L..15L} suggest that the magnetic dipole 
moment of light DM particle could not be larger than $1.6 \times 10^{-17} {\rm e \; cm }$.

\smallskip
On the experimental side, the findings of  DM searches are a puzzle that is difficult to resolve~\citep{2015NCimC..38...28P}:
several experimental collaborations in  direct DM searches have found experimental hints that could be related
with dark matter detection: DAMA/LIBRA~\citep{2008EPJC...56..333B,2010EPJC...67...39B} and possibly
CoGeNT~\citep{2011PhRvL.107n1301A,2011PhRvL.106m1301A} observed an annual modulation,
CRESST~\citep{2012EPJC...72.1971A} and CDMSSi~\citep{2013PhRvL.111y1301A,2013PhRvD..88c1104A} show
	hints of an excess of events. These experiments seem to indicate 
	the existence of a DM candidate with a mass of $\sim 10\;{\rm GeV}$ and a scattering 
	cross section on hydrogen and other chemical elements varying between 
	$10^{-41}$ and $10^{-36}\;{\rm cm^2}$.  The specific value of the scattering cross section  
	is  strongly dependent on the DM model used to interpret the data~\citep[e.g.,][]{2011PhRvL.107n1301A,2011PhRvD..84b7301D}. 
	Presently, the null results constraints are  from CDMSGe~\citep{2011PhRvL.106m1302A}, XENON~\citep{2011PhRvL.107m1302A,2012PhRvL.109r1301A} and LUX~\citep{2014PhRvL.112i1303A}.      
	These experiments found no evidence for an interaction of DM with baryons for the cited mass and scattering 
	cross section range, at least in the case of a contact type of the DM-nucleus interaction models. 
	Nonetheless, there are new theoretical proposals that resolve
	the differences between the different experimental results, the most successfully being       
	the long-range DM-nucleus interactions.
	In these type of DM models, the interaction between DM and baryons is not contactlike,  but occurs through a light particle mediator~\citep[e.g.,][]{2011PhRvD..84k5002F,2012JCAP...08..010D,2014PhLB..728...45F,2013JCAP...10..019C}. The impact of such a DM particle in the Sun's interior can modify significantly its core structure~\citep{2014ApJ...795..162L}.

%%%%%%%%%%%%%%%%%%%%%%%%%%%%%%%%%%%%%%%%%%%%%%%%%%%%%%%%%%%%%%%%%%%%%%%%%%%%%
\section{Dark matter and the Sun}
%%%%%%%%%%%%%%%%%%%%%%%%%%%%%%%%%%%%%%%%%%%%%%%%%%%%%%%%%%%%%%%%%%%%%%%%%%%%%
\label{sec:DMS}

As is usually done in these studies, we consider that the Sun's evolution  
in a DM halo is identical to the SSM. Likewise, these solar models 
are required to reproduce  the current Sun observables such as radius and luminosity.  
Therefore, the  models to compute the impact of DM in the evolution of the Sun were obtained as follows: for each set of DM parameters,
we compute  a solar-calibrated model following the same procedure used to compute a SSM~\citep{1993ApJ...408..347T}, i.e., by automatically adjusting the helium abundance and the convection mixing length parameter until the total luminosity and the solar radius are within $10^{-5}$ of the present solar values. Typically, a calibrated DM solar model is obtained after a sequence of
10 to 20 intermediate models.

\smallskip  

As the accretion of DM by the star produces minor differences in the Sun's core structure 
and almost no effect in the stellar envelope, these solar models follow the same 
path as the SSM in the 
Hertzsprung-Russell diagram. 
For the solar model of reference, we choose a SSM
with a low-metallicity composition~\citep{2009ARA&A..47..481A}, usually referred to as low-Z metallicity SSM.  
This SSM was computed using  
an  updated version of the stellar evolution code {\sc cesam}~\cite{1997A&AS..124..597M}.
The code has up-to-date microscopic physics, and in particular uses the
nuclear reaction rates from the NACRE Compilation~\cite{2011RvMP...83..195A}. 
This SSM predicts solar neutrino fluxes and helioseismic data that are consistent with other
SSM models found in the literature~\citep[e.g.,][]{2011ApJ...743...24S,2016JPhCS.665a2078T}. In relation to the properties
of our standard solar model, this can be found in~\citet{2013MNRAS.435.2109L}. 

\smallskip  
In a DM halo, a star captures DM from the beginning of the
premain sequence until the present age  ($4.6\; {\rm Gyr}$). 
The efficiency of the star in accumulating DM in its core is regulated  
by  three leading processes: capture, annihilation and evaporation  of DM particles.

The total number of particles $N_{\chi}(t)$ that accumulates inside
	the Sun at a certain epoch is computed by solving the following  differential equation
	\begin{eqnarray}
	\frac{dN_{\chi}(t)}{dt}=\Gamma_c - \Gamma_a N_\chi(t)^2  -\Gamma_{e}N_{\chi}(t),
	\label{eq:DMNs}
	\end{eqnarray}
	where $\Gamma_c$, $\Gamma_a$ and $\Gamma_e$ are the capture, annihilation and evaporation rates.
	A detailed account about these quantities can be found in~\citet{1996PhR...267..195J} and~\citet{2005PhR...405..279B}:
	\smallskip 
	
	- $\Gamma_c$ determines the amount of DM particles captured by the star.
	This quantity, among others, depends on the radius and escape velocity 
	at each step of the star's evolution. Nevertheless, it is the scattering of the DM particles 
	with baryons which is the leading process in the capture rate. The  
	scattering cross section depends on the mass and spin of the baryon nuclei.
	As usual, the scattering cross sections of DM particles with nuclei  $\sigma_{\chi}$ can be either a spin-dependent or spin-independent cross section, that are represented by $\sigma_{\chi,SD}$ and $\sigma_{\chi,SI}$.
	For all of the chemical elements excluding hydrogen, 
	the interaction with a DM particle is of spin-independent type (coherent scattering),
	for which the scattering cross section
	scales as the fourth power of the baryon nucleus mass number~\citep[e.g.,][]{2005PhR...405..279B}. 
	For hydrogen, the spin-dependent  interaction (incoherent scattering) is also taken into account.
	In our code the $\Gamma_{c}$ expression is computed following the original expression of~\citet{1987ApJ...321..571G,2004JCAP...07..008G} as described in~\citet{2011PhRvD..83f3521L}.
	
	\smallskip 
	
	- $\Gamma_a$  depends on the annihilation cross section $\langle \sigma v \rangle_{\chi}$ of particles and antiparticles. 
	In the current sets of DM models, we are uniquely concerned with $\langle \sigma v \rangle_{\chi}\approx 0$
	as discussed in the previous section. A detailed account 
	about the differences 
	 between the s-wave and p-wave DM annihilation channels 
	can be found in~\citet{2012ApJ...757..130L}. 
	\smallskip 
	
	- $\Gamma_e$ determines the amount of particles that evaporates from the Sun.
	In our study we use an approximate expression computed 
	for sun-like stars by~\citet{2013JCAP...07..010B} from the original work of~\citet{1987NuPhB.283..681G}. 
	Nevertheless, this should not much affect our result as we restrict our analysis to
	DM particles with a mass above $4\;{\rm GeV }$, for which evaporation is not significant~\citep{1990ApJ...356..302G}. 
	
\smallskip  
	In this study, the focus is on the interaction of DM with chemical elements heavier than hydrogen.
	The impact related to the capture of DM by the scattering off
	hydrogen was previously studied by~\citet{2010PhRvL.105a1301F}, among others. 
	
\smallskip  
Our DM models, if not stated otherwise, have the following
properties: the DM particles in the halo follow a Maxwell-Boltzmann velocity distribution, with a thermal velocity $v_{\rm th}=270\; {\rm km/s}$; 
the density of the DM halo is equal to  $0.38 \; {\rm GeV cm^{-3}} $~\citep[e.g.,][]{1995ApJ...449L.123G}; and
the stellar velocity of the Sun is $v_{\star}=220\; {\rm km/s}$. The mass
of the DM particle $m_\chi$, and the spin-independent and dependent scattering cross sections  with baryons $\sigma_{\chi,SD}$ and $\sigma_{\chi,SI}$ were chosen to be in agreement with the current experimental bounds for light DM particles. In particular, the spin-dependent and  scattering cross section $\sigma_{\chi,SD}$ is equal to $10^{-46}$ ${\rm cm^2}$. 
It is worth noting that, unlike in  previous studies, we solve numerically equation (\ref{eq:DMNs}), the equation  that regulates the accumulation of DM inside the star~\citep{2014ApJ...780L..15L}.
	
\smallskip  
% III
The DM impact on the star at each stage of evolution
is determined mostly by $N_{\chi}(t)$, the  number of DM particles  
accumulated by the star. Once the DM particles are captured by the Sun, these drift towards the Sun's central region, providing the star with  a new energy transport mechanism, which then removes energy 
from the core towards the more external layers of the star.
The efficiency of this transport of energy depends mainly on the ratio between the mean free path of the DM particles through the solar plasma $l_\chi$, and the characteristic radius of the DM particles distribution in the core of the star $r_\chi$~\citep[e.g.,][]{
2002MNRAS.331..361L,2008PhRvD..78l3510T}. For most of the DM-nuclei scattering cross sections 
$\sigma_{\chi}$ ($\sigma_{SD,\chi}$ or $\sigma_{SI,\chi}$) considered here, in which $l_\chi \ge r_\chi$,  the energy transport by DM is nonlocal. On the other hand, for large values of DM-nuclei scattering cross sections, in which $l_\chi \le r_\chi$, the DM particles are in local thermal equilibrium with the baryons. This latter regime applies only to values of $\sigma_{\chi}$ which are not considered in this work ($\sigma_{\chi}\ge 10^{-33}\, cm^2$).
However, we follow the prescription described in \citet{1990ApJ...352..654G} that extends the formalism developed for the local thermal equilibrium to other regimes by the use of tabulated suppression factors.   Moreover, once the characteristic radius of the DM core decreases with the mass of the DM particle, such as $r_\chi\propto m_\chi^{-1/2}$~\citep[e.g.,][]{2002MNRAS.331..361L}: stellar models computed for DM particles with different masses will produce 
the $^8B$ solar neutrino spectra with different shapes.

\smallskip  
The main effect of this additional transport of energy is  a decrease of temperature in the core of the Sun in relation to the standard solar model. This temperature variation is followed by an increase in the radial profiles of the density $\rho(r)$, and the mean molecular weight per electron $\mu_e(r)$. But as the increase of the density dominates over the increase of mean molecular weight per electron, and  the electron density $n_e(r)$ 
is proportional to the ratio $\rho(r)/\mu_e(r)$,  this  leads to an overall increase of $n_e(r)$  at core of the star~\citep[e.g.,][]{1983psen.book.....C}. 
Moreover, as the MSW effect (i.e., the conversion of electron-neutrinos to other neutrino flavors) increases with $n_e(r)$, this process leads to a decrease of survival probability of electron-neutrino, as it will be discussed in Sec.~\ref{sec:DMS8B}.   
Furthermore,  as the proton-proton chain and carbon-nitrogen-oxygen cycle of nuclear reactions are much more sensitive to the local variations of temperature than density, for a nuclear reaction  such as the $^8B$ $\beta$-decay process in the $^8$B nuclear reaction rate, this temperature reduction necessarily leads to smaller $^8B$ solar neutrino flux.

%%%%%%%%%%%%%%%%%%%%%%%%%%%%%%%
%%%%%%%%%%%%%%%%%%%%%%%%%%%%%%%
%%%%%%%%%%%%%%%%%%%%%%%%%%%%%%%

%%%%%%%%%%%%%%%%%%%%%%%%%%%%%%%%%%%%%%%%%%%%%%%%%%%%%%%%%%%%%%%%%%%%%%%%%%%%%
\section{$^8$B solar electron-neutrino spectrum and flavor oscillations}
%%%%%%%%%%%%%%%%%%%%%%%%%%%%%%%%%%%%%%%%%%%%%%%%%%%%%%%%%%%%%%%%%%%%%%%%%%%%%

\begin{figure}
	\centering
	\includegraphics[scale=0.5]{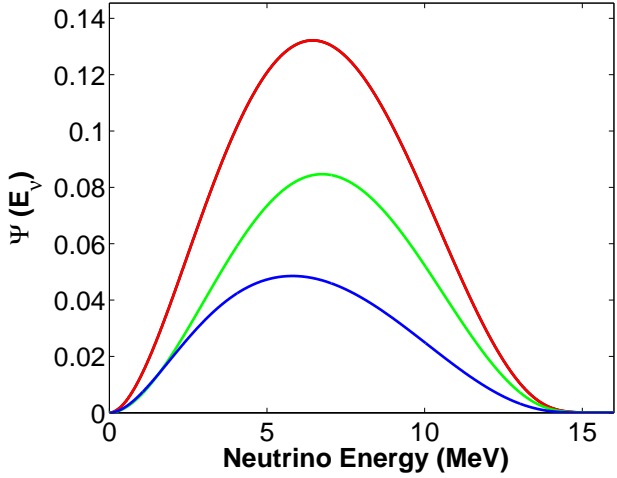}
	\caption{The $^8$B$\nu_e$ solar spectrum: 
		$\Psi^{e}_{\odot} (E_{\nu})$ -- electron-neutrino spectrum emitted in the Sun's core (red continuous curve); $\Psi^{e}_{\oplus} (E_{\nu})$ -- electron-neutrino measured by neutrino detectors on Earth 
		(blue continuous curve);   $\Psi^{\mu\tau}_{\oplus} (E_{\nu})$ -- nonelectron-neutrino spectrum
		(combine $\tau$ and $\mu$ neutrino spectrum) on Earth (green continuous curve). In the figure $\Psi^{e}_{\cdots} (E_\nu)$ corresponds to the probability per MeV that a electron-neutrino is emitted with a energy $E_\nu$.
		Notice that $\Psi^e_\odot (E_\nu)=\Psi^e_\oplus (E_\nu)+\Psi^{\mu\tau}_\oplus (E_\nu)$.
		This calculation used an up-to-date SSM (see text).}
	\label{fig:1}
\end{figure}

The $^8$B$\nu_e$ spectrum emitted by the $^8$B reaction in the Sun's core 
has been shown to be equivalent to several experimental determinations of the 
$^8$B$\nu_e$ spectrum~\citep[e.g.][]{2000PhRvL..85.2909O,2006PhRvC..73b5503W}. \citet{1986PhRvC..33.2121B,1987PhRvC..36..298N} among others have shown that 
the $^8$B$\nu_e$ neutrino spectrum emitted in the Sun's core is 
equal to the spectrum measured in the laboratory,
as the surrounding solar plasma does not affect this type of nuclear reaction.
Moreover, the $^8$B$\nu_e$  experimental spectrum agrees remarkably well with the  
theoretical prediction for neutrinos with an energy below 12 MeV.
In particular, the $^8$B$\nu_e$ neutrino spectra deduced from four laboratory experiments~\citep{2003PhRvL..91y2501W,2006PhRvC..73b5503W,2006PhRvC..73e5802B,2011PhRvC..83f5802K,2012PhRvL.108p2502R} agrees within about 1\% at high neutrino energies,  whereas before they differed by 4\%~\citep{2012PhRvL.108p2502R}.
Figure~\ref{fig:1}  shows $\Psi^{e}_\odot (E_\nu)$,
the $^8$B$\nu_e$ spectrum emitted 
by the $^8$B solar reaction in the Sun's core, with 
$\Psi^{e}_\oplus (E_\nu)$ and  $\Psi^{\mu\tau}_\oplus (E_\nu)$,
the
two components of the  $^8$B$\nu_e$ neutrino spectrum measured  
on Earth.
We note that the $\Psi^{e}_{\odot}(E_\nu)$ spectrum (cf. Fig.~\ref{fig:1})
is identical to the $^8$B$\nu_e$  spectrum measured in the laboratory. 
Therefore, the only variation expected in
the electron-neutrino spectrum measured by solar neutrino detectors, 
i.e., $\Psi^{e}_{\oplus}(E_\nu)$,   
is uniquely related to the neutrino flavor oscillations.

\begin{figure}
	\centering{\includegraphics[scale=0.45]{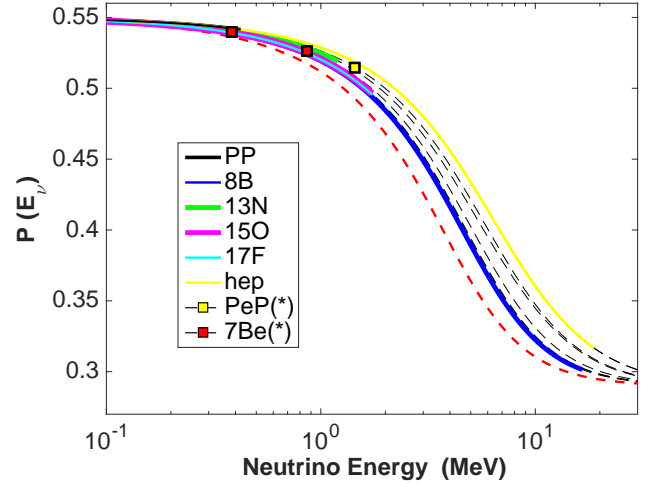}}
	\caption{The survival probability of electron-neutrinos as a function of the neutrino energy for a standard solar model.  The colored parts of the curves indicates the energy range of neutrinos  produced in the Sun's core for each nuclear reaction (as "measured" by solar neutrino experiments):  $^8$B$\nu_e$ (blue curve), $^7Be\nu_e$ (two red-squares; emission lines),  
		$pep\nu_e$ (yellow square, emission line),
		$hep\nu_e$ (yellow curve)   $pp\nu_e$ (black curve),  $^{13}N\nu_e$ (green curve),  $^{15}O\nu_e$ (magenta curve), $^{17}F\nu_e$ (cyan curve).
		The reference curve (red dashed curve) defines the survival probability of electron-neutrinos in the center of the Sun.  The generic properties of such curves can be found in~\citet{2013PhRvD..88d5006L}.}
	\label{fig:2}
\end{figure}

The fraction of electron-neutrinos that changes flavor 
depends on the parameters associated with vacuum and matter oscillations, 
and this latter process depends also on the local properties of the solar plasma~\citep{2013PhRvD..88d5006L}.
This is the reason why $\Psi^{e}_\oplus(E_\nu)$ is significantly different from $\Psi^{e}_\odot(E_\nu)$.
These quantities are related as follows:
\begin{eqnarray} 
\Psi^{e}_\oplus(E_\nu)=\langle P_{\nu_e} (E_\nu)\rangle \Psi^{e}_\odot(E_\nu)
\label{eq-Psie}
\end{eqnarray}
where $\langle P_{\nu_e} (E_\nu)\rangle$ is the electron-neutrino survival probability
of a neutrino of energy $E_\nu$. $\langle P_{\nu_e} (E_\nu)\rangle$ reads  
\begin{eqnarray} 
\langle P_{\nu_e} (E_\nu)\rangle = 
A^{-1} \int_0^{R_\odot} P_{\nu_e} (E_\nu,r)\Phi_\nu (r) 4\pi \rho(r) r^2 dr, 
\label{eq-A}
\end{eqnarray}  
where  $\Phi_\nu (r)$ is the $^8$B electron-neutrino emission source. 
As usual, $r$ is the solar radius, $\rho(r)$ is the density and $A$ is a normalization constant.
In the absence of matter-induced oscillations due to the Earth's atmosphere,
$P_{\nu_e}(E_\nu, r)$ corresponds to the electron-neutrino survival probabilities 
on Earth during the day. It follows that  
$P_{\nu_e}(E_\nu, r) = \cos^4{\theta_{13}}\;P_{2\nu_e}(E_\nu, r)+\sin^4{\theta_{13}}$, 
where $P_{2\nu_e}(E_\nu, r) $ is the probability of a two-flavor neutrino oscillation  
model~\citep[e.g.,  ][]{2011PhRvD..83e2002G,2010LNP...817.....B,1932ZPhy...78..847L}
and $\theta_{13}$ a neutrino mixing angle in vacuum. $P_{2\nu_e}(E_\nu, r) $ is
given by
\begin{eqnarray} 
P_{2\nu_e}(E_\nu, r) =  \frac {1} {2} + \frac {1} {2}  \cos {(2 \theta_{21})} \cos{(2 \theta_m)},
\label{eq-B}
\end{eqnarray} 
where $\Delta m_{12}$ is the mass  difference between two flavors,
$\theta_{21}$ is a flavor mixing angle in vacuum and
$\theta_{m}$ is the matter mixing angle inside the Sun.
$\theta_{m}$ reads
\begin{eqnarray} 
\sin{(2\theta_m)}= 
\frac{\sin{(2\theta_{12})}}{\sqrt{(V_m-\cos{(2\theta_{12})})^2+ \sin^2{(2\theta_{12})}}},
\label{eq-C}  
\end{eqnarray}
where $ V_m (E_\nu,r) = 2 \sqrt{2} G_f \; n_e (r)\; E_\nu\;\cos^2{(\theta_{13})}/\Delta m_{21}$,
$ G_f $ is the Fermi constant and $ n_e (r) $ is the electron density of the solar plasma.
Equations~(\ref{eq-A}-\ref{eq-C}) determine the probability of electron-neutrinos to be converted 
to other flavors when propagating in matter.
This process can affect all solar neutrino sources, but  it is more pronounced on the  $^8B\nu_e$ spectrum.

\smallskip 

Figure~\ref{fig:2} illustrates this specific point. In the figure it is shown the "theoretical" (dashed black curves)   and the "observable" (colored  curves) survival probability $\langle P_{\nu_e}(E_\nu)\rangle $ of electron-neutrinos~\footnote{The "theoretical" $\langle P_{\nu_e}(E_\nu)\rangle $ although not directly related with the real solar spectrum unlike the  "observable" $\langle P_{\nu_e}(E_\nu)\rangle $. This quantity illustrates well the effect that the energy dependence of neutrino matter oscillations have on the flux of electron-neutrinos.} as a function of neutrino energy for the SSM. $\langle P_{\nu_e}(E_\nu)\rangle $ was computed  for  $^8$B$\nu_e$,  as well as for other neutrino source
reactions of the proton-proton chain and carbon-nitrogen-oxygen cycle.
Although all neutrino's nuclear reaction sources occur in the 
	Sun's core, the only ones that can be affected the structure changes due to accretion DM in the Sun's core are the ones that produce the neutrinos with
	the higher energy. This corresponds to the  $^8$B neutrinos (blue curve) and
	hep  neutrinos (yellow curve),  as shown in Figure~\ref{fig:2}. Nevertheless, the former occur near the center of the Sun and are measured with much better precision that the hep neutrino spectrum.
	Therefore, the $^8B\nu_e$ spectrum will be the most affected by the presence
	of the DM in the Sun's core.
\begin{figure}[ht]
%	\centering
\subfigure{\includegraphics[scale=0.45]{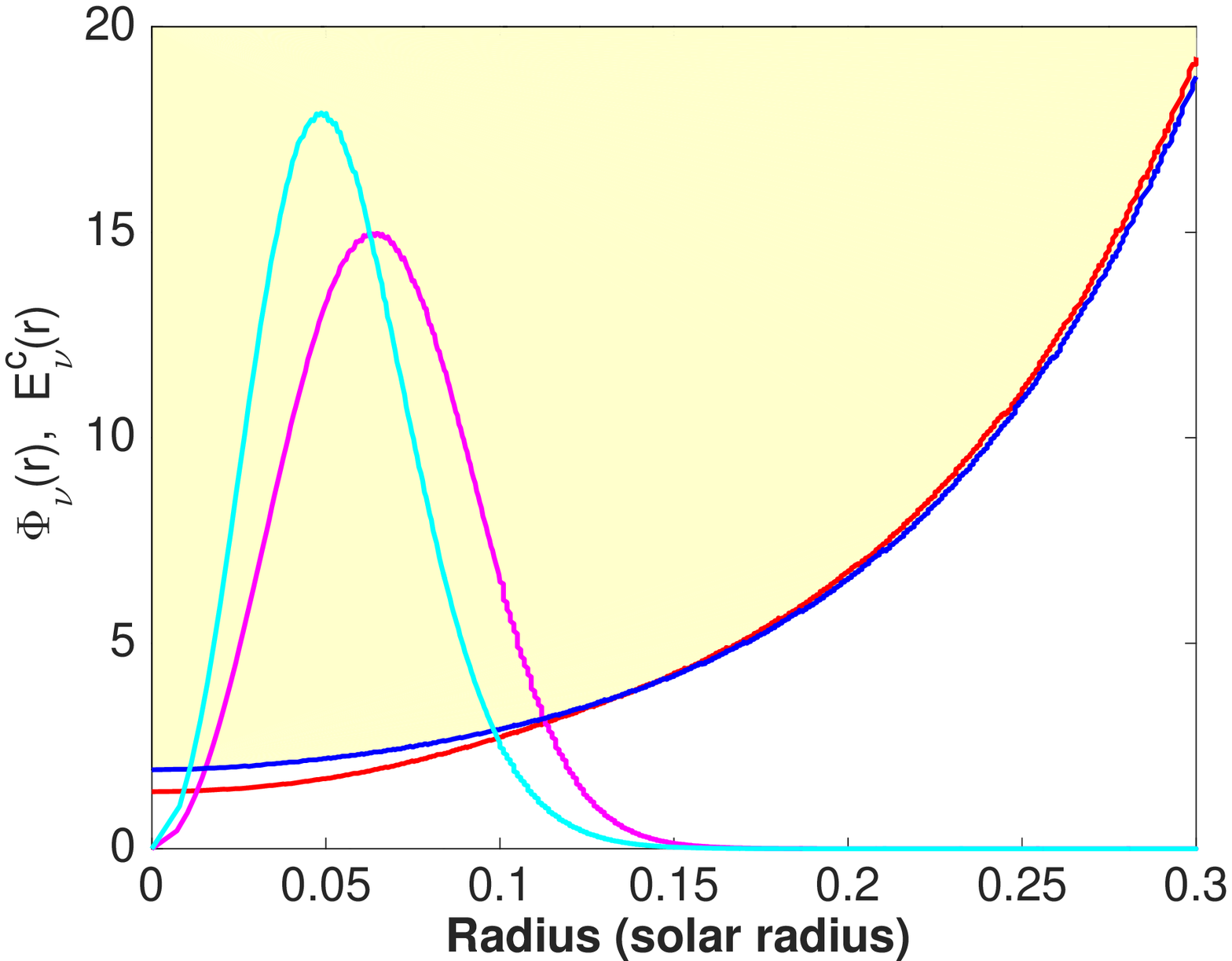} }
\caption{Variation of the functions $\Phi_\nu(r)$  and
E$_\nu^{\rm c}(r)$ (in MeV)  with the fractional solar radius $r$. 
The cyan and blue curves correspond to $\Phi_\nu(r)$  and E$_\nu^{\rm c}(r)$ for the SSM, and the magenta and red curves are the equivalent ones for a DM solar model with the $m_\chi =5\; GeV$ and $\sigma_{\chi,SI}=10^{-36}\;{\rm  cm^2} $. Both sets of curves have relatively similar shapes. {\bf Cyan and magenta curves:} 
 $\Phi_\nu$'s is drawn as a function of $r$ such that $\Phi_\nu(r)=F^{-1} df(r)/dr$ for which $f(r)$ is the $^8B$ neutrino flux in $s^{-1}$ and $F$ is the total neutrino flux for $^8B$ nuclear reaction rate. {\bf Blue and red curves:}  E$_\nu^{\rm c}$'s   represented as a function of $r$ 
 corresponds to the minimum neutrino energy E$_\nu$ that a neutrino must have in order to experience a resonance (see text). Accordingly, electron neutrinos such as $E_\nu \ge  E_\nu^{\rm c}(r)$ will experience matter flavor oscillations in the solar core, otherwise this effect is negligible.
}\label{fig:3}
\end{figure}

The impact of DM on the  $\langle P_{\nu_e}(E_\nu)\rangle $  or
$^8B\nu_e$ spectrum can be described as follows:
In the Sun's interior, a neutrino of energy $E_\nu$ can be converted to 
other flavors if  $E_\nu\ge E_\nu^{\rm c}(r) $. 
The quantity $E_\nu^{\rm c}(r) $ defines the minimum (critical) energy that  
a neutrino must have to be strongly affected by flavor oscillations.  
E$_\nu^{\rm c}(r)$ is determined by the condition  
$V_m(E_{\nu}^{\rm c},r)=\cos{(2\theta_{12})}$ (from Eq.~\ref{eq-C}), it follows that
$E_\nu^{\rm c}(r)= \Delta m_{21} /(2 \sqrt{2} G_f)\;  
\cos{(2\theta_{12})}/\cos^2{(\theta_{13})} \; n_e^{-1} (r) $.

\smallskip
The survival probabilities of electron-neutrinos and $E_\nu^{\rm c}(r) $ were 
computed by using the fundamental parameters of solar neutrino oscillations in the vacuum:
$\Delta m_{12}$ and $\theta_{12}$ as determined by the KamLAND experiment~\citep{2011PhRvD..83e2002G}. 
Although the contribution related to $\theta_{13}$ is very small, 
we take its contribution into  account by choosing $\theta_{13}=9\; {\rm deg}$, 
a value that is in agreement with current experimental measurements~\citep{2012PhRvD..86a3012F,2012PhRvD..86g3012F}.
Figure~\ref{fig:3} shows the critical value $E_\nu^{\rm c}(r) $ for current SSM and other solar models: 
neutrinos experiment MSW flavor oscillations in regions of the Sun's core
where the neutrino energy is such that $ E_\nu \ge  E_\nu^{c}$ (yellow region in Fig.~\ref{fig:3}), 
otherwise the effect is insignificant. 
The magnitude of  flavor oscillations caused by matter depends 
on  the local value of $ n_e (r) $, namely the values of density and metallicity. 
These oscillations are only significant in the Sun's core and negligible in most of the radiative 
region and solar convection zone. The fraction of electron-neutrinos converted to other flavors 
depends also on the location of the neutrino source, as well as the local temperature as shown 
in Fig.~\ref{fig:3}.

\begin{figure}
\centering{\includegraphics[scale=0.45]{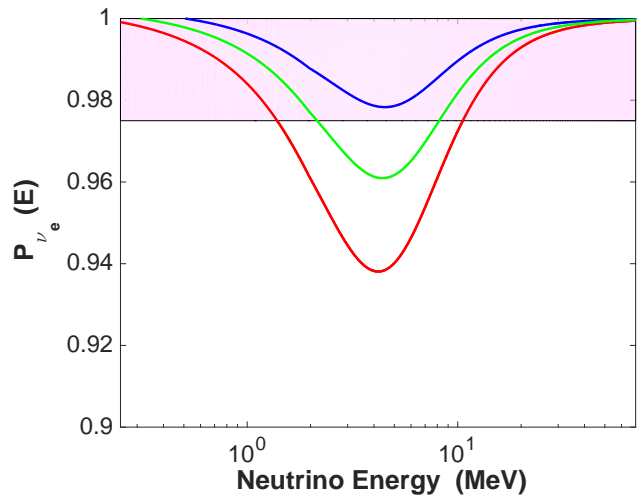}}
\caption{The ratio of several $^8B$ survival probabilities of electron-neutrinos of DM solar models in relation to the standard solar model.
The colour curves correspond to an halo of DM particles
with the following properties: $m_\chi=4 \; {\rm GeV}$ and $\sigma_{\chi,SI}=10^{-37}\;{\rm cm^2}$ ({\bf red curve});  
$m_\chi=5 \; {\rm GeV}$ and $\sigma_{\chi,SI}=10^{-35}\;{\rm cm^2}$ ({\bf green curve}) and
$m_\chi=7 \; {\rm GeV}$ and $\sigma_{\chi,SI}=10^{-35}\;{\rm cm^2}$
({\bf blue curve}). The pink area defines the experimental error bar of
the LENA detector (see text). }
\label{fig:4}
\end{figure}

\begin{figure}
\centering{\includegraphics[scale=0.45]{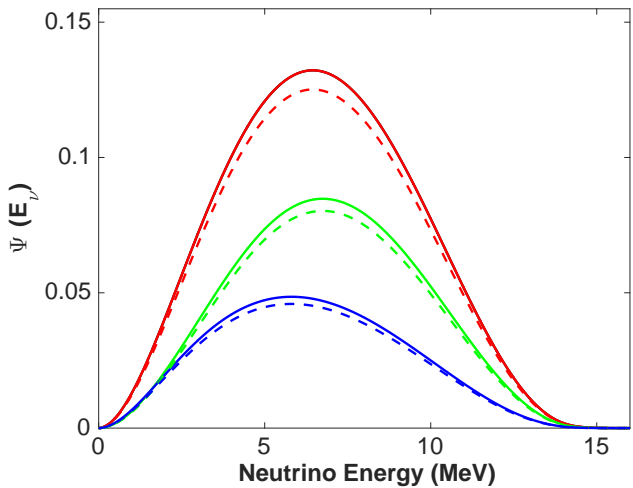}}
\caption{
The $^8$B$\nu_e$ solar spectrum: the continuous corresponds to the SSM and  the dashed curve to DM solar model with $m_\chi =5\; GeV$ an $\sigma_{\chi,SI}=10^{-36}\;{\rm  cm^2} $.
The colour scheme is the same as the one used in figure~\ref{fig:1}.}
	\label{fig:5}
\end{figure}

%%%%%%%%%%%%%%%%%%%%%%%%%%%%%%%%%%%%%%%%%%%%%%%%%%%%%%%%%%%%%%%%%%%%%%%%%%%%%
\section{Dark matter signature on $^8$B neutrino spectrum}
%%%%%%%%%%%%%%%%%%%%%%%%%%%%%%%%%%%%%%%%%%%%%%%%%%%%%%%%%%%%%%%%%%%%%%%%%%%%%
\label{sec:DMS8B}

The presence of DM in the Sun's core changes its thermodynamic structure, 
modifying the temperature, density and chemical composition, 
as well as $n_e(r)$. Although the effect is relatively small,
as neutrinos are very sensitive to the temperature 
of the Sun's core, minor variations in temperature produce  variations 
in the $^8$B$\nu_e$ spectrum. Consequently, the $^8$B neutrino flux and $^8$B$\nu_e$ spectrum are modified as a result of the variation of the magnitude 
and location of the $^8$B neutrino source (cf. Fig.~\ref{fig:3}).
In addition, the variation of $n_e(r)$ distorts the $^8$B$\nu_e$ spectrum, 
due to an alteration of the survival probability of electron-neutrinos
which determines the fraction of electron-neutrinos converted to other flavors. Different DM models  have different critical neutrino energies $E_{\nu}^{\rm c}$, leading to distinct  $\langle P_{\nu_e} (E_\nu)\rangle$ for $^8$B$\nu_e$ and other neutrino sources (cf. Figs~\ref{fig:2},~\ref{fig:3} and~\ref{fig:4}).
The combination of these different physical processes modifies the shape 
of $\Psi^{e}_\oplus(E_\nu)$, i.e., the $^8$B$\nu_e$ spectrum measured in terrestrial  detectors.

\medskip
% ILIDIO      
	The  $^8$B$\nu_e$ spectrum  is strongly dependent on the temperature, but also on the density
	and chemical composition.  Actually, $\Psi^e_\oplus(E_\nu)$ the $^8$B$\nu_e$ spectrum shape of the electron neutrino, is related to the variation of the density by three possibilities: the production rate of electron-neutrinos leading to the neutrino function  $\Phi_\nu(r)$, the location of the maximum of $\Phi_\nu(r)$ and the survival probability of electron neutrinos (i.e., the conversion of electron-neutrinos to other flavors, Eq.~\ref{eq-A}):      
	
	- The first two effects result from the fact that variations in total neutrino flux $\phi$ (or equally on the production rate of $^8$B neutrinos) depend on the temperature T and density $\rho$ as $\Delta \phi/\phi\approx \Delta \rho/\rho+\alpha \Delta T/T$ where $\alpha=24.5$ is obtained from~\citet{1993ApJ...408..347T}. Accordingly, a $10\%$ variation on $\Delta \phi/\phi$ is either attributed to a variation in $10\%$ of density, $0.4\%$ in temperature, or a combination of both. Moreover, a similar
	variation on the molecular weight is also expected.  Nevertheless, as mentioned by several authors~\citep[e.,g.][]{2002PhRvD..66e3005B,2010PhRvL.105a1301F,2010PhRvD..82j3503C}
	as solar models are calibrated to have the observed solar radius and luminosity, the effective variation of $\rho$ is smaller than the previous estimate, the mitigation coming from the temperature and chemical composition readjustment.  The variation of  $\Phi_\nu(r)$ leads to a slight change in the location of the maximum of  $\Phi_\nu(r)$.  This variation also influences the amount of electron-neutrinos converted to other flavors as described by equation (\ref{eq-A}). 
	
	- Equally from Eq.~(\ref{eq-B}), the variation of $\Delta P_e/P_e$ is proportional to the variation of electronic density $\Delta n_e/n_e$ (or density and molecular weight). The impact of DM on the electronic-neutrino survival probability function $ \langle P_{\nu_e} (E_\nu)\rangle $ is shown in Fig.~\ref{fig:4}.  As pointed out by previous authors~\citep[e.g.,][and references therein]{2013PhRvD..88d5006L} the effect on $ \langle P_{\nu_e} (E_\nu)\rangle $ at first order is relatively small, as at low energies the neutrino oscillations are vacuum-related and therefore insensitive to the Sun's structure; for the higher energy neutrinos, the flavor oscillations are vacuum and density-related (see Fig.~\ref{fig:3}). The effect of the Sun's structure on $ \langle P_{\nu_e} (E_\nu)\rangle $ is more pronounced for neutrinos with intermediate energies (from 0.1 to 1.0 MeV). As shown in Fig.~\ref{fig:4}, the variation of  electronic density (density and molecular weight) with solar radius slightly changes the profile of $ \langle P_{\nu_e} (E_\nu)\rangle $, leading to small changes in the shape of the $^8$B $\nu_e$ spectrum (see Sec.~\ref{sec:DMS}).
	Moreover, Fig.~\ref{fig:4} shows the part of the $^8$B$\nu_e$ spectrum that is more affected. This corresponds to neutrinos with an energy in the range: 1 to 10 MeV. This variation is more pronounced  
	for light DM particles with the largest scattering cross sections. This effect
	 reduces  the $^8$B$\nu_e$ electron survival probability curve by as much as 6\% in relation to the standard case. It is important to observe that such effect on the electron survival probability will distort the electron-neutrino $^8$B spectrum in the same neutrino energy range.
	 Although this shape deformation is small, once  future measurements of $^8$B electron neutrinos will be able to possibly detect such types of effect, if observed it could provide a hint of the existence of dark matter. It is worth remembering that the shape of the $^8$B neutrino spectrum is very well measured by current laboratory experiments (see introduction and references therein). For DM solar models discussed in this paper, the maximum effect observed in
	$\Psi^{e}_\oplus(E_\nu)$ uniquely related with $\langle P_{\nu_e} (E_\nu)\rangle $ 
	is of the order of 6.5\% and occurs near  6 MeV.              

\medskip
The identification by a future solar neutrino detector of a strong distortion in the shape of $\Psi^{e}_\oplus(E_\nu)$ compared to that predicted by the SSM, 
would  constitute  a strong hint for the presence of DM in the Sun's core. 
The magnitude of the distortion should give some indication about the amount of DM  and the extension of the DM core. Figure~\ref{fig:5} shows the difference for the  $^8$B$\nu_e$ spectrum for  a DM solar model. 
This is due to the fact that $\nu_e$ neutrinos of different energy have a different sensitivity to the local distribution of electron density of the Sun's core, specifically, only the more energetic neutrinos are affected by matter flavor oscillations. 

\medskip
In this study we have explored how the presence of DM in the Sun's core
changes the shape of solar neutrino spectra, for instance the $^8$B$\nu_e$ neutrino
spectrum. In many cases, the impact of  DM in the Sun's core can be determined by variations on the total neutrino fluxes due to local temperature changes.   Nonetheless, there is an important point to make: even for an identical percentage variation $\Psi^e_\oplus(E_\nu)$ and $\phi$, there is a fundamental different between both quantities, as the former gives the location where  the DM effect occurs (cf. Fig.~\ref{fig:5}).  Indeed, the presence of DM in the solar core will distort $\Psi^e_\oplus(E_\nu)$ very likely around  $E_\nu\sim$ 6 MeV (although depending on the DM models, $\Psi^e_\oplus (E_\nu)$ could be quite singular) and not uniformly distributed, information that is not possible
to obtain from $\phi$. As $\phi$ is an integrated quantity $\Phi_\nu(r)$, this only give us very limited information about the radial distribution of the DM in the core.

Although  this work is focused on studying the $^8$B$\nu_e$ spectrum, 
our study is easily extended  to other neutrino sources as shown in Fig.~\ref{fig:2}.
However,    there are currently no neutrino experiments planned to measure the spectrum of other solar neutrino sources 
in the near future.

%%%%%%%%%%%%%%%%%%%%%%%%%%%%%%%%%%%%%%%%%%%%%%%%%%%%%%%%%%%%%%%%%%%%%%%%%%%%%
\section{Summary}
%%%%%%%%%%%%%%%%%%%%%%%%%%%%%%%%%%%%%%%%%%%%%%%%%%%%%%%%%%%%%%%%%%%%%%%%%%%%%

%2016PhRvD..94f3512M

In this paper, we have shown that a detailed measurement of the $^8$B$\nu_e$  spectrum in the range from 1-15 MeV by future solar neutrino experiments will permit us to probe in great detail the core of the Sun (below $0.1 R_\odot$) in  a search for traces of DM. We have also shown that this type of DM diagnostic can be extended  to other solar neutrino spectra, once the experimental data becomes available (cf. Fig.~\ref{fig:2}).

The SSM predicts that the $^8$B$\nu_e$ earth spectrum, expected to be measured by solar neutrino detectors, is very different from the $^8$B$\nu_e$ solar emitted spectrum,  due to vacuum and matter oscillations which neutrinos experience when travelling to Earth. The presence of DM in the Sun's core will change the magnitude and shape of the $^8$B$\nu_e$  spectrum for terrestrial observers in a very distinct manner. 
Since there are many astrophysics  processes that are not yet included in the standard solar model, that could also affect the physics of Sun's core and the solar neutrino fluxes~\citep[e.g.,][]{Lopes2018},
 the distortion of the $^8$B$\nu$ spectrum is an additional important
signature  that could play a determinant role in disentangling the impact of DM
from other possible physical processes. The next generation of solar
neutrino detectors like JUNO~\citep{2016JPhG...43c0401A} and LENA~\citep{2012APh....35..685W} should be able to 
achieve the required precision to test such solar DM models. 
This will be achieved  by simultaneously increasing the precision
on the measurements of  the solar neutrino spectrum (or the survival
electron-neutrino probability) and also by increasing the 
energy resolution, without which it is not possible to precisely measure
the shape distortion of the $^8$B $\nu_e$ spectrum.
Moreover, it is expected that the LENA detector after only 5 years of  measurements will be able to obtain a probability  survival for
electron-neutrino (or the equivalent  $^8$B$\nu$ flux) 
with an experimental  error smaller than 0.025~\citep{2014PhLB..737..251M}.
This precision is sufficient to put constrains in some solar DM models~\citep{2010Sci...330..462L}, since for some of them the survival electron-neutrino probability variation is of the order of 0.06 (cf. Fig.~\ref{fig:4}). For instance, if we assume that such
experimental accuracy is attained  on LENA measurements,
solar DM models with $m_\chi\le 5 {\rm GeV} $ and $\sigma_{\chi, SI}\ge 10^{-35} {\rm cm^2}$ as the ones shown in Figure~\ref{fig:4}
can be excluded using the putative LENA data set.
This type of diagnostic could help to determine if light asymmetric DM is indeed present in the Sun's core, as this type of DM has been suggested as a nonstandard solution to resolve the solar abundance problem.

\bigskip
% % % % % % % % % % % % % % % % % % % % % % % % % % % % % % % % % % % % % % % % % %
\begin{acknowledgments}
The authors are grateful to the anonymous referee for the comments and suggestions which improved the overall quality this article.
The author I.L. thanks the Funda\c c\~ao para a Ci\^encia e Tecnologia (FCT), Portugal, for the financial support
to the Center for Astrophysics and Gravitation (CENTRA/IST/ULisboa) 
through the Grant No. UID/FIS/00099/2013. The research of J.S. has been supported at IAP by  the ERC project  267117 (DARK) hosted by Universit\'e Pierre et Marie Curie - Paris 6  and at JHU by NSF grant OIA-1124403.
We are grateful to the authors of the DarkSUSY and CESAM codes for having made their codes publicly available.
\end{acknowledgments}
% 
%\bibliographystyle{yahapj}
%\bibliography{mn8Blib}
% % % % % % % % % % % % % % % % % % % % % % % % % % % % % % % % % % % % % % % % % % % % % % % % %
% % % % % % % % % % % % % % % % % % % % % % % % % % % % % % % % % % % % % % % % % %

\end{document}